\documentclass[runningheads]{svmult}

\usepackage{makeidx}   
\usepackage{graphicx}  
\usepackage{subeqnar}  
\usepackage{multicol}  
\usepackage{physprbb}  
\makeindex             

\begin{document}

\title*{Evidence for a Supernova in the Ic Band Light Curve
of the Optical Transient of GRB 970508}

\toctitle{Evidence for a Supernova in the Ic Band Light Curve
of the Optical Transient of GRB 970508}

\titlerunning{Evidence for a Supernova}

\author{V. V. Sokolov}

\authorrunning{V. V. Sokolov}

\institute{Special Astrophysical Observatory of RAS,
	   Karachai-Cherkessia, Nizhnij Arkhyz, 369167 Russia}

\maketitle

\begin{abstract}
Unique data on $BVRI$ brightness curves of the OT of GRB 970508
obtained with the 6-m telescope have been interpreted in the framework of the
GRB-SN (supernovae) connection. The effect must be maximal in the $I_c$ band
as OT GRB 970228 \cite{Galama1998}. The peak absolute ($M_B$) magnitude of
the suggested SN must be around -19.5 for the OT of GRB 970508.
If all or the main part of long GRBs are associated with SNe, the GRB host
galaxies (for ground-based observations, at least) must
be dimmer than the peak magnitude of the SN \cite{Sokolov2001}).
\end{abstract}

\begin{figure}[t]
\centering
\includegraphics[width=0.75\textwidth,bb=50 40 535 420,clip]{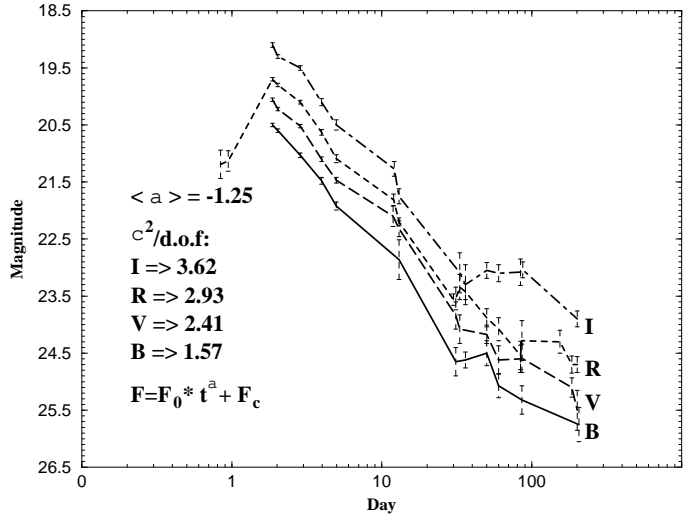}
\caption{The maximum of the effect is in $I$-band and decreases toward the
blue part of spectrum as it should be for SN Ib/c. The data from
the SAO-6m telescope (May 1997 -- Aug 1998) were used.
\cite{Sokolov1998}, \cite{Sokolov1999}.
The deviation from an "average" power-law is shown.
}
\end{figure}

The $I_c$ band light curve of GRB 970508 afterglow exhibits a rebrightening
phase, or a "shoulder" \cite{Sokolov1998}. This spectral property lasts for
not less than 51 days \cite{Sokolov1999}.
Over this period, the color index V-I increases by 1.6 mag,
while $I_c$ = 23.04+/-0.14 for the OT. The deviation from the "average"
power-law exceeds 1 magnitude.
Recent papers show that GRB optical afterglows have unusual temporal and
$VRI$-spectral properties: GRB 970228 \cite{Galama1998},
GRB 980326 \cite{Bloom99a} , and possibly GRB 990712,
and GRB 991208 (Castro-Tirado A., in these proceedings).
All these OTs have temporal and spectral characteristics
similar to those of GRB 970508, for which the $I_c$ band light curve is
consistent with the assumption that the SN emission overtakes the OT flux at
late times, i.e. nearly 31 days after the GRB. It should be noted that the
emission of all type-I SNe shows a strong UV deficit owing to absorption
lines below 3900\AA. Thus, a weak flux is expected from the suggested SN in
the $R_c$ band for a redshift $z$ of 0.835 corresponding to the $U_{rest}$
(3652\AA) in the rest frame. In contrast, the SN brightness must be
maximal at larger wavelengths (Fig1),
i.e. in the $I_c$ band as observed for GRB 970508
(and for GRB 970228 with $z = 0.695$ \cite{Galama1998}), plus a time
profile stretching for the light curve of the SN by a factor of $1+z$.
As $z = 0.835$, the $I_c$ band
in the observer frame corresponds to the $B_{rest}$ (4448\AA)
band in the rest frame,
the observed $I_c$ flux is thus used to derive the peak absolute magnitude
($M_B$) of the assumed SN. At the "shoulder"
the brightness of GRB 970508 OT corresponds to the maximum
luminosity of a peculiar SN with $M_B$ around -19.5
if for GRB990708 host we have $M_{B_{rest}}$ around -18.5
\cite{Sokolov1999} (Fig.2).

We can consider the type Ib/c SNe as
a preliminary model of GRB/SN gamma-ray bursters, but the peak magnitudes
of Type Ib/c (or core collapse) SNe are not constant: $M_B$ is
from -16 to -19.5.
Thus, the GRB host galaxies (for ground-based observations,
at least) must be dimmer than the peak magnitude of the SN (see Table 1 in
\cite{Sokolov2001}).
SN1997ef has been recognized as a peculiar SN Ic from its light curve
(Mazzali P., et al., in these proceedings).
It shows a very broad/flat peak and a slow tail.
But the luminosity is lower than that of SN 1998bw. However, for the purposes
of this report, the shape of SN Ic light curve is sufficient as an example
of Type Ib - Ic SN.
We do not know exactly the light curve (and the spectrum)
of the true SN accompanying GRBs; may be in the OT GRB 970508
case it was a peculiar SN by its luminosity and by the shape of
the light curve (Fig.2).
\begin{figure}[t]
\centering
\includegraphics[width=0.75\textwidth,bb=30 30 720 522,clip]{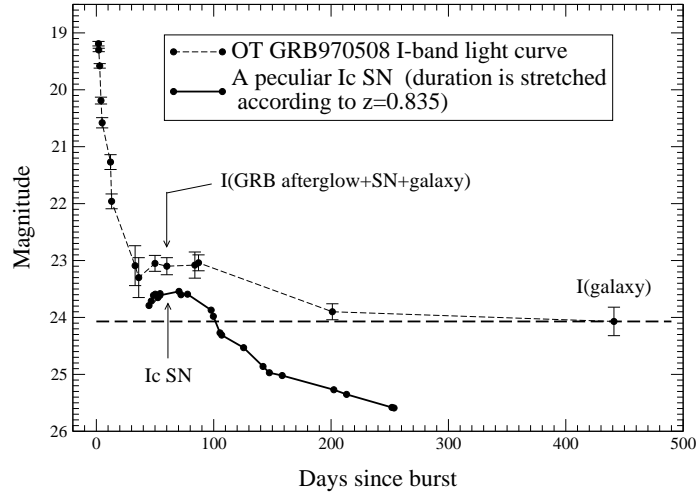}
\caption{OT GRB~970508 $I$-band light curve plus a kind of a pecular Ic SN
(by the shape of the light curve) similar to SN 1997ef, but with the
peak luminosity like SN 1998bw.}
\end{figure}

Let us suppose that all or at least the main part of long
{\bf GRBs are associated with SNe} (but not all SNe are associated with GRBs).
From the observations of 1997-2000 for known GRB host galaxies we have:
$m_{HostGal}$ is from 22 to 28.5 mag \cite{Bloom2000}.
It leads to the fact that
the search for direct GRB-SN associations in galaxies close to us
is a challenge, because the majority of the SNe related to GRBs
will be faint (22-26 mag) and in very distant galaxies with $z\ge 0.4-4.5$.
For a number of galaxies brighter than 26 mag
in one square degree of the sky we have
$N_{gal < 26 mag} \approx 2\cdot 10^{5}\cdot gal\cdot deg^{-2}$
\cite{Casertano2000}.
Taking into account the observations with BATSE
$ n_{GRB} \approx 0.01\cdot deg^{-2} yr^{-1} $,
we conclude that
the rate of GRB events is
$N_{GRB} \approx 0.5\cdot 10^{-7}\cdot yr^{-1} galaxy^{-1}$.
The SNe rate for the local starburst galaxies is
$ N_{SN} \approx 0.02\cdot yr^{-1} galaxy^{-1}$
for  all types of SNe combined.
It should be kept in mind that GRB are not observed in all galaxies and
that not all types of SNe can be related to GRBs, but only Type Ib/c
\cite{Capellaro1999}.
So we have $ N_{SN} \approx 0.001\cdot yr^{-1} galaxy^{-1}$.
But if one looks at the matter the other way round, we must remember:
the massive star-forming in very distant (and dusty) GRB galaxies
with $z\sim > 1$
is more vigorous than in local starburst galaxies or than in galaxies
like the Milky Way \cite{Blain2000},
giving rise of SN bursts to 10 times.
Thus, with due regard for these reservations, an estimate for
the number of GRBs which could be related to SNe
is of order of $N_{GRB}/N_{SN}\approx 5\cdot (10^{-5} - 10^{-6})$.
So, if some GRB -- SN relation really exists, then we have two possibilities:
{\bf 1)} Either we deal with a very very rare type of SNe (hypernova) --
GRBs related only to the $10^{-5}-10^{-6}$-th part of all observed SNe
in distant galaxies (up to 28th mag).
{\bf 2)} Or we have a very strong {\bf $\gamma$-ray beaming} with a solid
angle up to
$\Omega_{beam} \approx 5\cdot (10^{-5} - 10^{-6})\cdot 4\pi$
if GRBs are associated with an {\bf asymmetric} SN explosion
and the $\gamma$-ray beam
is directed towards an observer on Earth.

Thus, if all (or the main part) of long GRBs are associated with SNe,
we need more examples to test the GRB-SN link. But from Earth we can see
only peculiar SNe (by their luminosities \& by light curves) connected with
GRBs. The SNe must be brighter than their hosts. The luminosity of these dusty
galaxies $M_B$ are between about -18.6 and -21.3 \cite{Sokolov2001}.
{\it Acknowledgements:} This work was supported by INTAS N96-0315,
"Astronomy" Foundation (grant 97/1.2.6.4) and RFBR N98-02-16542.

\end{document}